# Fake News and Phishing Detection Using a Machine Learning Trained Expert System

Benjamin Fitzpatrick*
Department of Electrical and Computer Engineering
University of Alabama
3043 H.M. Comer, 245 7th Avenue
Tuscaloosa, AL 35487
Phone: +1 (205) 348-6351
Fax: +1 (205) 348-6959

Xinyu "Sherwin" Liang*
Dallas College – North Lake
5001 N. MacArthur Blvd.
Irving, TX 75038
Phone: +1 (972) 273-3000

Jeremy Straub                              *** Corresponding Author ***
Institute for Cyber Security Education and Research
North Dakota State University
1320 Albrecht Blvd., Room 258
Fargo, ND 58108
Phone: +1 (701) 231-8196
Fax: +1 (701) 231-8255
Email: jeremy.straub@ndsu.edu

* These authors contributed equally to this work

**Abstract**

Expert systems have been used to enable computers to make recommendations and decisions. This paper presents the use of a machine learning trained expert system (MLES) for phishing site detection and fake news detection. Both topics share a similar goal: to design a rule-fact network that allows a computer to make explainable decisions like domain experts in each respective area. The phishing website detection study uses a MLES to detect potential phishing websites by analyzing site properties (like URL length and expiration time). The fake news detection study uses a MLES rule-fact network to gauge news story truthfulness based on factors such as emotion, the speaker's political affiliation status, and job. The two studies use different MLES network implementations, which are presented and compared herein.  The fake news study utilized a more linear design while the phishing project utilized a more complex connection structure. Both networks' inputs are based on commonly available data sets.

# 1. Introduction

Artificial intelligence (AI) techniques allow computer software to make recommendations or decisions. Some characterize this as mimicking a form of human intelligence; however, in many cases computers are able to exceed human capabilities in terms of speed, accuracy and other metrics. AI techniques are typically used, in programming, to allow computer software to perform tasks not possible with traditional if-then, case, loop and similar structures. AI technique use, thus, allows for processes formerly performed by humans, such as translation or image recognition, to be performed instead computationally.

One of the most common forms of artificial intelligence is neural networks [1]. Neural networks approximate the neuron structure of the human brain. Artificial neurons (called nodes) are interconnected connected and have weight values and activation functions, allowing for complex and responsive networks to be generated.

Neural networks are part of a sub-discipline of computer and data science known as machine learning. Machine learning techniques are used to learn behavior via training and to discover specific patterns from large amounts of data. Several different forms of machine learning techniques exist. Common techniques include support vector machines [2], naive Bayesian classifiers [3], decision trees [4], and K-nearest neighbors [5] algorithms, which are often used for classification. Other techniques, such as linear regression [6] and support vector regression [2] are commonly used for data regression. Supervised learning, which neural networks fall under, is a task-driven approach while unsupervised learning is a data-driven approach. Common unsupervised learning techniques are K-means [7], K-medoids [8], and hidden Markov model [9].

While machine learning algorithms have been used in numerous areas, they also have weaknesses. One large issue with many AI and machine learning techniques is the problem of bias. This bias problem has led Noble to refer to some machine learning algorithms as "algorithms of oppression" [10] due to their potential impact on historically disenfranchised groups. AI bias and discrimination have prospectively been manifested towards many groups of people, especially the disadvantaged. Examples include discrimination against ethnic minorities [11] and the poor [12]. While a critical issue, bias is not the only issue presented by AI and machine learning. Doyle contends that the misuse of big data is, in some cases, so damaging that it goes as far as to threaten democracy [13].

This paper reports on the development and optimization of expert system networks for two studies. Both applications, phishing website detection and fake news detection, are key areas of national concern. With fake news classification, for example, there is a trade-off between the need to protect the public from intentionally deceptive content and preventing censorship. The fact that Twitter flagged numerous tweets as fake news during the 2020 presidential election caused some Twitter users to concern as they believed that Twitter was limiting free speech [14]. Phishing prevention, similarly, pits warning users about a potential danger against similar censorship concerns.

Explainable and defensible AI systems may offer some benefit in this circumstance, as they could explain what the cause for a flagging or warning action was. This would allow internet users to analyze and make decisions for themselves.

For both studies, the Machine Learning Expert System (MLES), introduced in [15], was used to develop a rule-fact network representation which is reported on herein. The input fact weighting values within these rules will be refined over time, using gradient descent, to optimize outputs as described in [15]. The goal of both studies is to make decisions that are accurate and based on only causal relationships and to be able to explain the results so that users can understand why a particular decision (e.g., flagging a tweet or warning about a phishing website) was made.

This paper explores the applicability of MLES and the different ways it can be used as part of the two studies. It continues, in Section 2, with a review of prior work which provides a foundation for the work presented herein. Section 3 presents the phishing website detection study and the MLES rule-fact network that was developed for it. Section 4 presents the fake news detection study and the MLES network that it utilizes. Section 5 compares and analyzes these two studies. Finally, Section 6 discusses key conclusions from this work and potential areas of future work.

**2. Background**

This section presents work in three areas that provides a foundation for the work presented in this paper. First, prior work on artificial intelligence and machine learning is presented. Then, prior work on expert systems is reviewed. Finally, the machine learning trained expert system, the system used for the work presented herein, is discussed.

*2.1. Artificial Intelligence and Machine Learning*

Artificial intelligence (AI) is used throughout modern society. They provide a wide variety of benefits [16]. The technologies have been used to play games [17], find software bugs [18], help the handicapped [19], detect hackers [20] and command robots [21]. Machine learning (ML) systems adaptively learn on their own or utilize an explicit training process. Rewards and supplied input and output data [22] can be used to direct learning [23]. Other systems are left to find data associations on their own [24].

For supervised training system, gradient descent and backpropagation [25] techniques are used to apply changes throughout a neural network or similar structure. To do this, they use an iterative process which is based on identifying and correcting differences between network-under-training output and target output values. A variety of customizations to these basic techniques have been suggesting including those that optimize speed [26], focus on bias elimination [27], [28] and attack resilience [29], [30]. Other approaches utilize other AI techniques, such as genetic algorithms [31] and particle swarm optimization [32] as part of the training process.

*2.2. Expert Systems*

Expert systems, which use rule-fact networks for inference [33], were first developed in the 1960s and 1970s [34], [35]. In expert systems' most basic form, facts can have binary values and rules identify facts that can be asserted (marked true), based on other facts already being true. They use a rule processing engine to scan the collection of rules and facts, called the system's knowledge base, for these rules with satisfied preconditions. Expert systems which use fuzzy logic concepts [36] have been proposed. These systems can represent the nuance or uncertainty in fact values and rules. Newer forms of expert systems use other techniques [37], beyond basic rule-fact networks, including genetic algorithm and neural network concepts.

Perhaps the most advanced form of expert system, prior to the work presented in [15], was proposed by Mitra and Pal [38] who defined a so-called "knowledge-based connectionist expert system." This system was to start with "crude rules", which would be stored as connection weights in a neural network. Neural network training could then be used to refine the ruleset. This system does not appear to have been implemented and it is unclear how it would prevent the network structure from being trained away in the neural network.

Expert systems have been used across a variety of fields. Examples include power systems [39], agriculture [40], education [41] and medicine [42], [43]. Fuzzy logic expert systems, in particular, have been utilized for autonomous vehicle obstacle avoidance [44], medical applications [45]–[47] and even software architecture evaluation [48].

## *2.3. Machine Learning Trained Expert Systems*

This work uses a defensible machine learning-trained expert system approach, that was introduced in [15]. This system combines aspects of expert systems and neural networks and utilizes a gradient descent training technique. Unlike systems proposed prior to this, that technique is performed directly on the expert system's rule-fact network. While there is a limited conceptual similarity to the suggestion in [38] of using neural networks for expert system rule development and refinement, the approach proposed in [15] does not allow the training process to modify the structure, preventing the system from learning illogical, invalid and problematic associations. Instead, the training process is used to distribute error-correction to rules' input fact weightings.

The system incorporates the concepts of partial membership and ambiguity, allowing fact values to be between 0 and 1. Rules, instead of just being logical 'AND' operators, store weighting values that define the contribution of each input fact towards the identified output fact. These weighting values must be between 0 and 1 and sum to 1. As the typical training algorithms used with neural networks, which have different and more rigid structures are not suitable for expert systems, a training algorithm for expert system networks was proposed in [15].

The operations of the system are presented in Figure 1 and the algorithm that is used for determining the level of change that needs to be applied is presented in Figure 2.

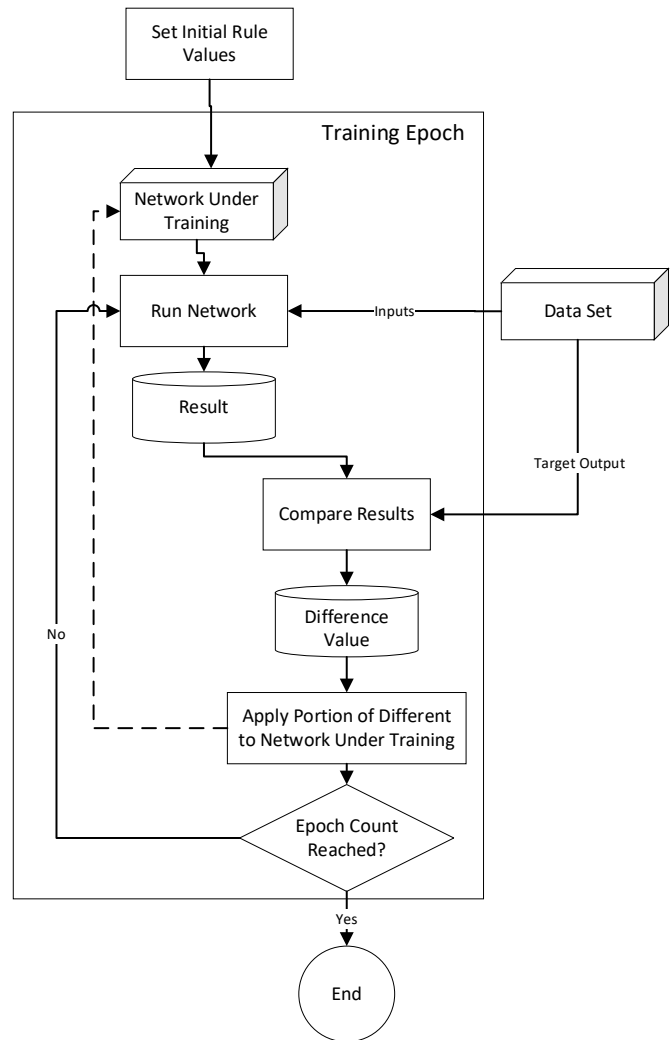

Figure 1. Training Process (modified from [15]).

This system identifies rules that contribute to the output fact. During training, the system-under-training is compared to the supplied data and a portion of the difference is applied to each contributing rule. The exact level of change that is made is based on a configurable velocity value. In [15], the system was trained using a 'perfect' system training technique, as described in [49]; however, in this work it is trained using data from two datasets in a manner that is typical of training in most neural network studies.

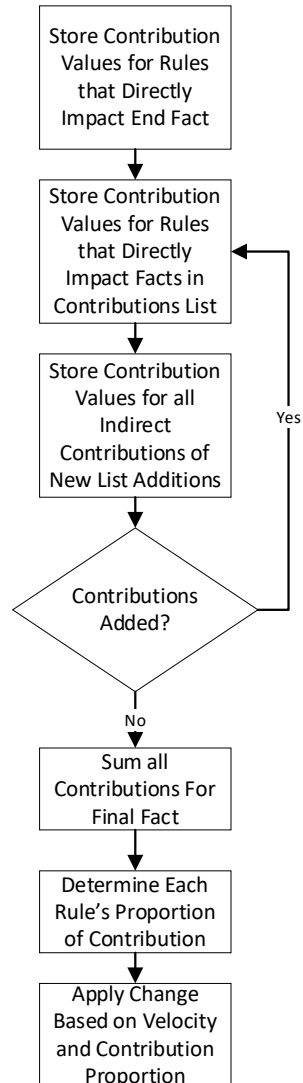

Figure 2. Algorithm for determining node change levels [15].

## 3. Phishing Website Detection Study

This study explores the detection of phishing websites based on features of the URL and the website it is associated with. The section continues by providing a brief overview of phishing is. It then discusses the dataset that the MLES network has been designed based on and the information it contains. The processing needed to prepare it for use by the MLES system is also discussed. Finally, in this section, the MLES network design is reviewed, with a focus on its key features and the overall methodology used for its development.

### 3.1. Topic Overview

Phishing [50], [51] is a commonly used social engineering attack. It is typically used to gain information that can be used for authentication or to impersonate a user. By sending messages through email and other mediums, which in many cases impersonate a friend, co-worker or trusted entity, an attacker can obtain credentials or other confidential information without having to breach any physical or digital system. In this form of attack, the human involved is the 'system' that has to be breached. The attackers, thus, are using psychological and other tricks to attempt to get the target to give them the information that they desire.

Several techniques exist for identifying phishing attacks. Both "blacklists"1 and machine learning techniques are used for this purpose. For blocklists, a list of known-bad URLs and domains is created and consulted by message filtering software to determine if a message that has been received is likely to be a phishing attack. Several sources for these lists, such as PhishTank [52], exist. Some provide user-submitted URLs whereas others blocklists are built autonomously using machine learning models [53].

When machine learning is used, multiple characteristics of a message are analyzed. This includes scanning for known phishing keywords and phrases. The makeup of the target URL and the properties of the website that it links to can also be considered. A neural network can be used to make a determination regarding the legitimacy of the message and its contents.

### *3.2. Dataset*

A dataset developed by Vrbančič [54] was selected for use in this study. It was originally created for evaluating the effectiveness of using deep neural networks for phishing detection. The dataset includes 88,647 entries and 111 features per entry. These features are divided into categories based on the different portions of the URL (as shown in Figure3). The dataset also includes general information about external features of the domain.

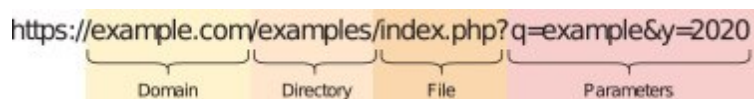

Figure 3. Sections of a URL [55].

### *3.2.1. Data Preprocessing*

Preprocessing was performed to this data to decrease the number of features that would serve as network inputs to the MLES. First, symbol characteristics were combined to form category-wide symbol classifications. Additionally, the values from the dataset were scaled to fall within the range of zero to one used by the MLES software processing system. Additionally, as some entries were missing data for certain features, such as if the URL did not include any parameters or directory, entries were modified so that unreported features were set to the average value of this feature in the dataset, to ensure that the missing feature is not erroneously treated as a zero value.

The dataset was divided into training, testing, and validating subsets. This ensured that the data records that were used to test the effectiveness of the system were not used in the training of the system.

### *3.3. Network Design*

The network implemented to characterize this study for processing in the MLES system (see Figure 4) is largely divided based on the categories that make up the dataset. These are based on the components of the URL (and there is a category for other website information). In most cases, input facts serve as input facts to more than one rule. This allows an increased number of comparisons to take place, as opposed to a more linear design (such as the design used for the fake news detection study which is described subsequently). This study has taken the approach of creating rules to interconnect and create intermediate facts[2] for all logically associated input facts. While this matrix-style design (shown in Figures 5 to 9) is used at low levels, each area is then encapsulated into a single fact that serves as the interface node to other areas of the network. This encapsulating node is at the bottom of each figure, in Figures 5 to 9. These encapsulating facts then participate in rules which results in a final prediction regarding a given presented URL being a phishing attempt website, given the feature data that is presented.

---
[1] The term "blacklist" is used pervasively throughout the computing and cybersecurity literature to refer to this technique. A more appropriate and inoffensive term is "blocklist". This term will be used throughout the remainder of this paper.

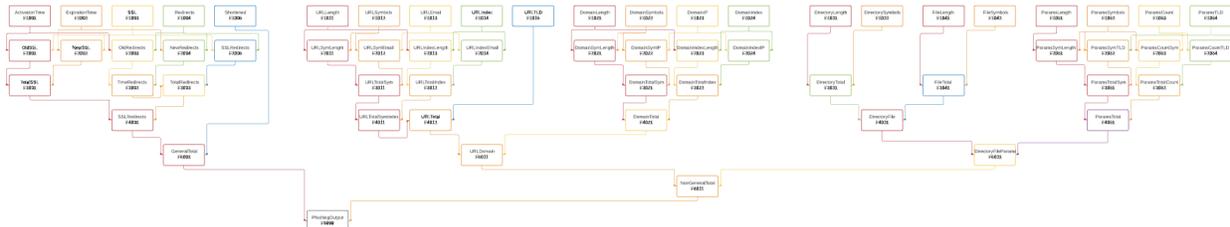

Figure 4. High level overview of the rule-fact network.

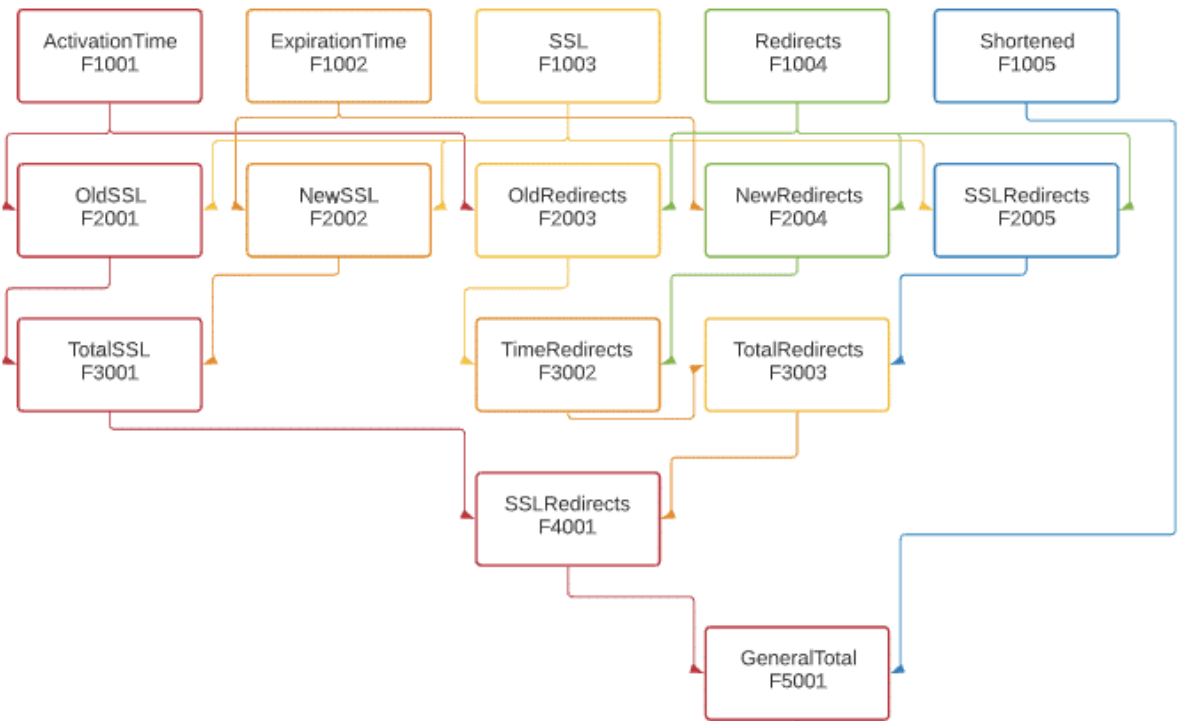

Figure 5. Rule-fact network (rules not shown for simplicity) for the general category facts.

---

[2] The term intermediate fact denotes a fact that doesn't correspond to a specific discrete piece of information but rather is used to represent a complex structure within the MLES system.

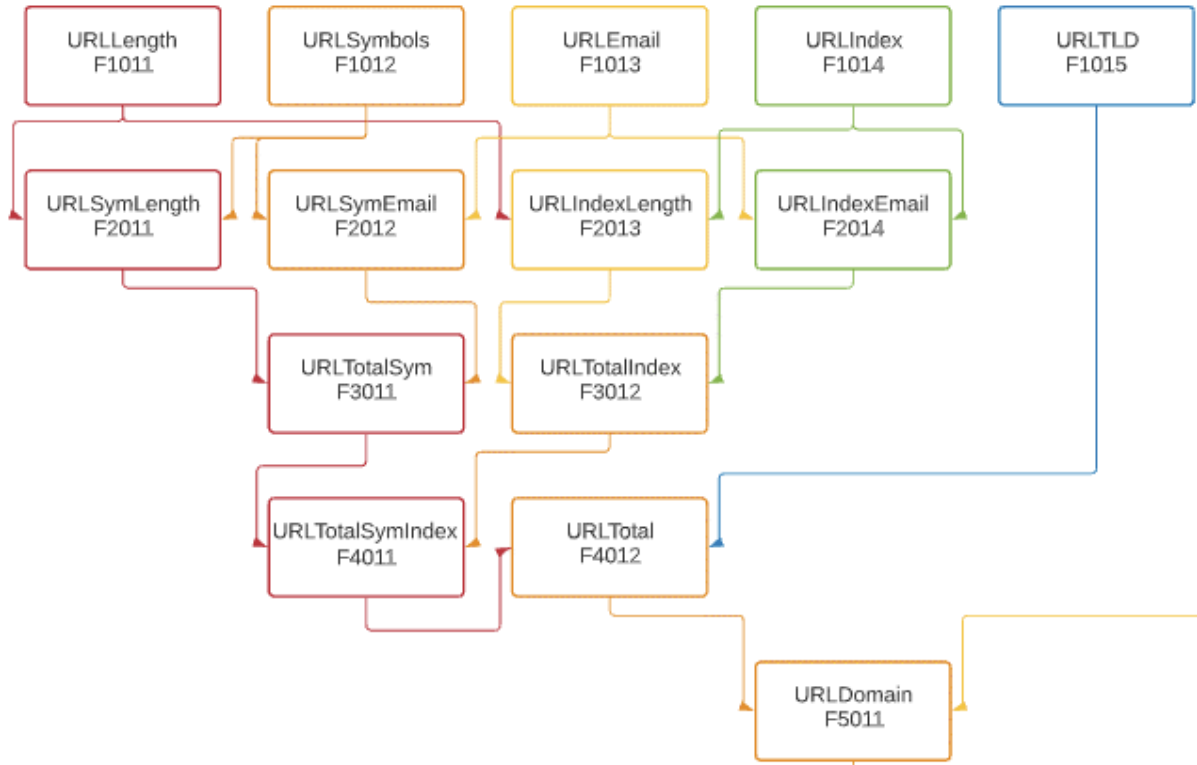

Figure 6. Rule-fact network (rules not shown for simplicity) for entire URL related facts.

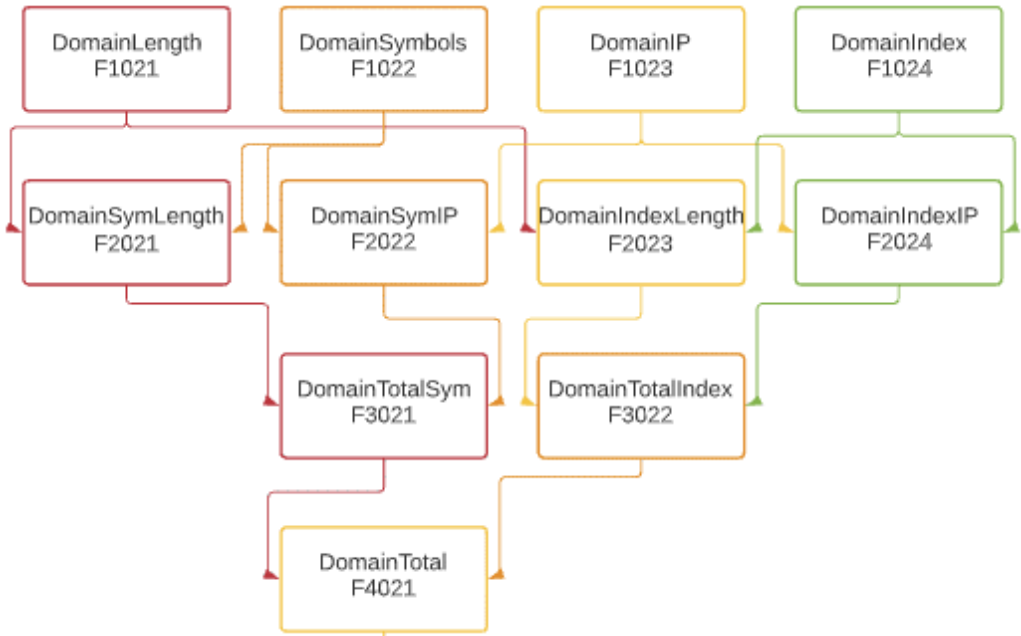

Figure 7. Rule-fact network (rules not shown for simplicity) for domain-related facts.

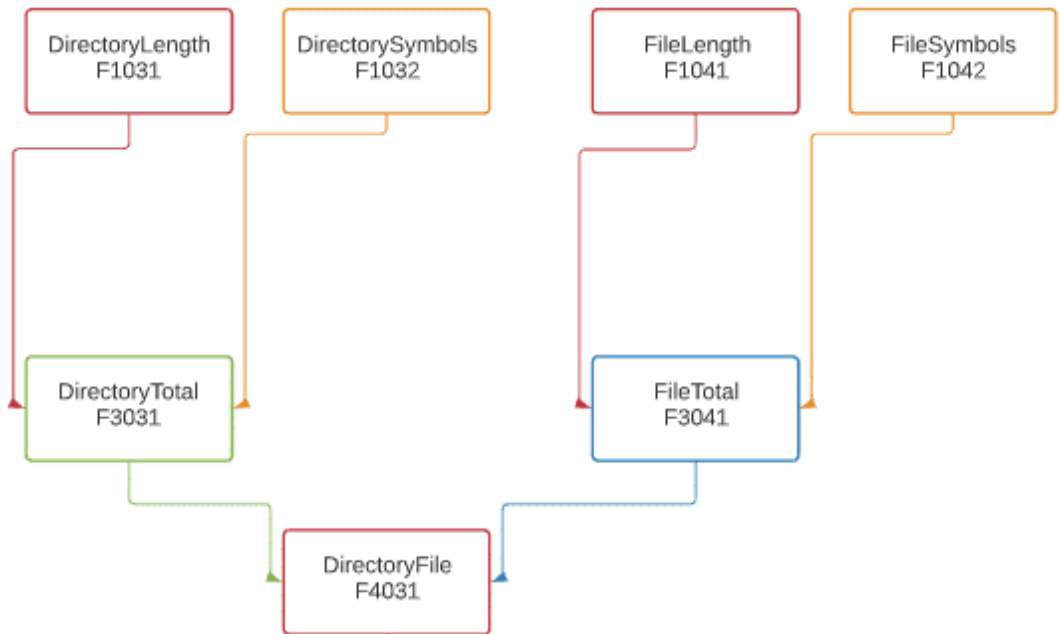

Figure 8. Rule-fact network (rules not shown for simplicity) for the directory (path) and file URL components.

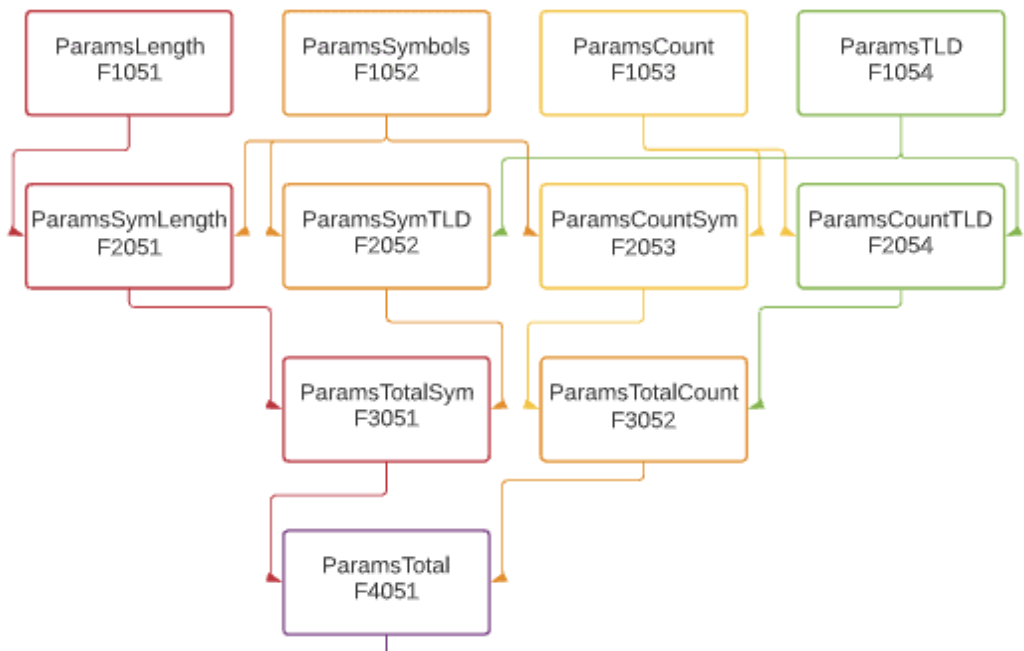

Figure 9. Rule-fact network (rules not shown for simplicity) for the parameters component of the URL.

There is not a single correct way, which is known at present, to design rule-fact networks for use with the MLES system. For each application, the designer has to make decisions and identify facts that correlate with the desired output and determine what facts should be grouped together and encapsulated. Additionally, designers have to make a fundamental decision regarding the level of network interconnectedness and how interconnected a network should be. A key focus of this paper is the comparison of the two design approaches used by the two studies herein. Identifying optimization techniques and heuristics for MLES networks will be a key area of future work.

At present, an approach of making and refining decisions based on system performance after training is required. This facilitates optimization and improvement beyond what is provided by the machine learning training system in MLES which solely modifies rule weightings.

**4. Fake News Detection Study**

This section provides a description of how the MLES is used for the fake news detection study. It begins with a brief overview of the dangers of fake news to society. This is followed by a description of ways that are utilized to identify fake news. Next, a discussion of how the MLES is used to build a human-understandable explainable and defensibly causal rule-fact network. The selection of the dataset, the necessary pre-processing of the data and the network construction process are all discussed.

*4.1. Danger of fake news*

The rise of social media and other factors have resulted in lots of online content of varying degrees of accuracy. Some of this content, commonly known as fake news, is deliberately deceptive. If believed and followed, it may represent a significant threat to society. Fake news has been blamed for armed attacks [56], influencing presidential elections [57] and numerous other maladies [58]. Individual consumers have been shown to be under-prepared for dealing with information which isn't pre-screened for them for accuracy [59], [60] and to place undue trust in online content [61]. In fact, 55% of Americans have indicated that they are getting news from social media [62], without the benefit of fact checking. Perhaps more concerning is that 75% of Americans have indicated that they have believed fake news headlines [63].

If fake news can be identified, a variety of approaches are possible for preventing its negative impact. These range from warnings and alerts [58] to other more restrictive approaches.

*4.2. Techniques to detect fake news*

A common method of identifying fake news is to do so manually. There are already several fact-checking websites such as FactCheck.org, AFP Fact Check USA, and PolitiFact that present manual fact checking of some news items. Of course, manual fact checking has an obvious disadvantage of being resource intensive and comparatively slow.

A variety of fake news identification techniques, which use machine learning technology, have been implemented and evaluated. Researchers have demonstrated the capability to identify fake news articles based on features such as the word choice used in text [64], rhetoric structure [65], and the sentiment revealed by the text [66]. Sentiment analysis, which is sometimes referred to as opinion mining, attempts to characterize the writer's view of the article's subject. It is conducted using techniques such as natural language processing. The result of this analysis is typically a number, where a positive value indicates a positive sentiment and a negative value indicates sentiment negativity.

A number of open-source systems for sentiment analysis have been developed. IBM, for example, provides a natural language processing service called Waston Natural Language Understanding for text analytics [67]. Upadhayay and Behzadan used this resource to analyze every sentence in the LIAR dataset and create emotion indexes. They added the results of this analysis to the dataset and named it Sentimental LIAR [66]. This dataset is used for this work.

*4.2. Dataset*

The Sentimental LIAR dataset [66] is utilized for this work. Sentimental LIAR is based on an earlier dataset called LIAR [68]. Compared to the LIAR dataset, the new dataset has five added sentiment fields: anger, fear, joy, disgust and sadness. Additionally, the developers of the new dataset have created a truthfulness Boolean value field, which is based on (but reduces the fidelity of) the more nuanced multi-level truthfulness classifications included in the original LIAR dataset. Five fields are also included which represent the "detected level of … emotional states" [66].

The dataset includes 30 fields and 12,786 records. The records are divided into three subsets: the train set with 10,236 records, the test set with 1,267 records and the validation set which has 1,283 records.

*4.3. Data processing*

It is necessary to process the dataset to facilitate its importation and use in the MLES. This processing is now discussed.

First, the text-based label field is transformed into a numerical value. The label field has six possible values: pants-fire, FALSE, half-true, barely-true, mostly-true, and TRUE. These were converted to the values of 0.0, 0.1, 0.5, 0.6, 0.75, and 1.0.

For free text data, typographical and spelling corrections were made. The general approach to this type of data started with counting the number of occurrences of each word or phrase across all entries in the training dataset. Those with too few occurrences (i.e., less than 20) were replaced with "other" or "unknown", as appropriate for the field. For example, compared to the Democratic and the Republican parties, the less frequently included Green party was be replaced with "unknown". From here, the text needs to be transformed into a number that the MLES system can process. This is done by determining what the average truth numeric score is for each word or phrase and utilizing that. For blank values, the average value of the particular field is used. As required by the MLES, all numbers were on a scale from 0 to 1. Notably, other approaches to this text processing step are possible. Further analysis in this area has been identified as a prospective topic for future work.

*4.4. Network design*

To facilitate analysis of the truthfulness of each news article, a MLES rule-fact network was developed. In developing this network, the role of each indicator in detecting fake news was analyzed. Indicators that related to the same area were grouped together. Unlike the phishing study, presented previously, facts were used in a single location in the network and were not connected as inputs to multiple rules, creating a linear network structure. In some cases intermediate facts (which don't correspond to a specific piece of knowledge) were utilized to facilitate the grouping of multiple input facts.

An overview of the network that was developed is presented in Figure 10. Figures 11 to 14 provide detailed views of each area of the network. The input facts were divided into three primary categories: text background (Figure 11), speaker background (Figure 12) and emotion scores (Figure 13). The text background includes the sentiment score for the communication and the type of communication that it is. The speaker background includes the speaker's state, party, job and a credibility score. Finally, the emotion scores are the five scores discussed previously based on textual analysis.

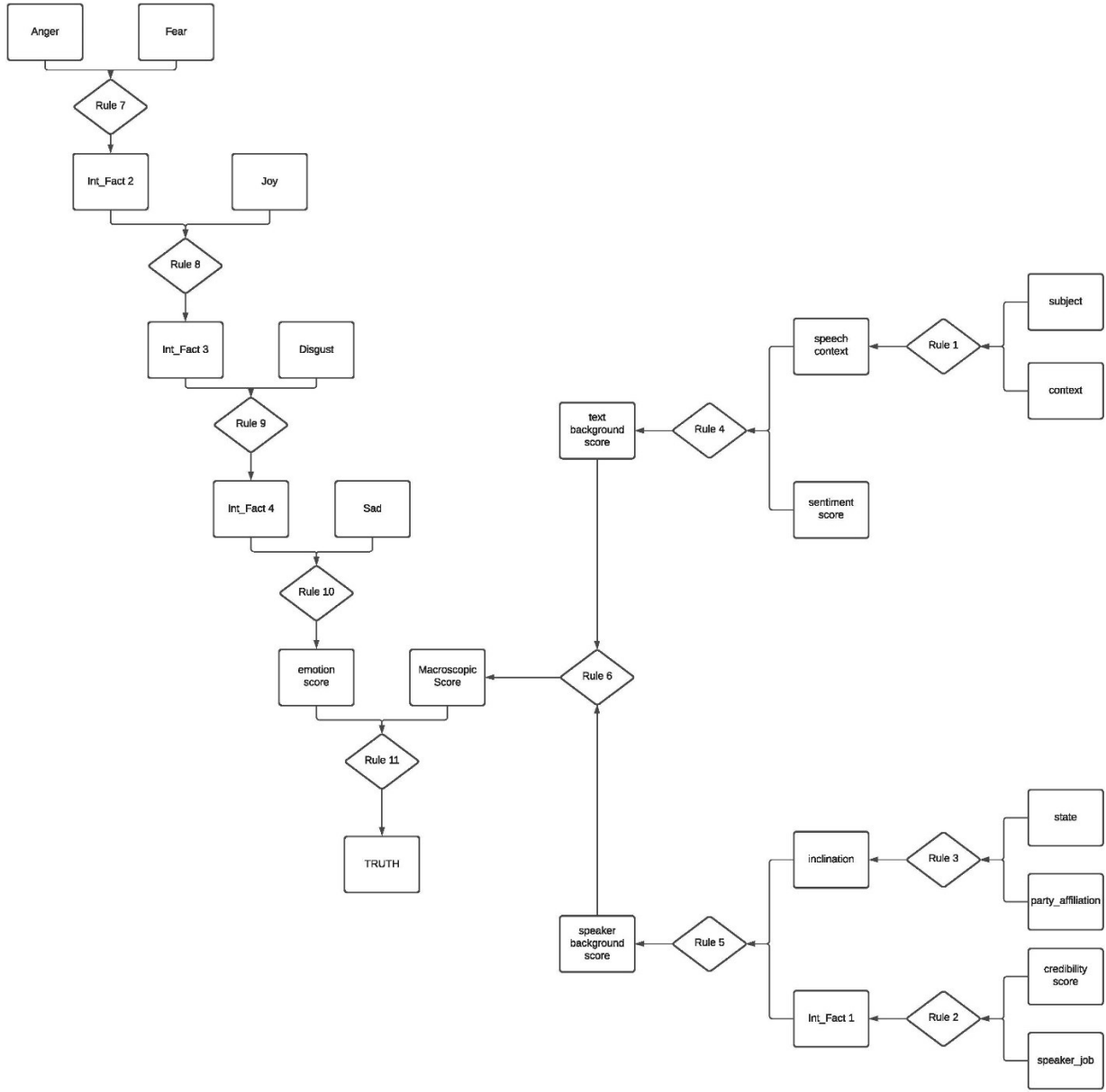

Figure 10. Overview of the rule-fact network used for the fake news detection study.

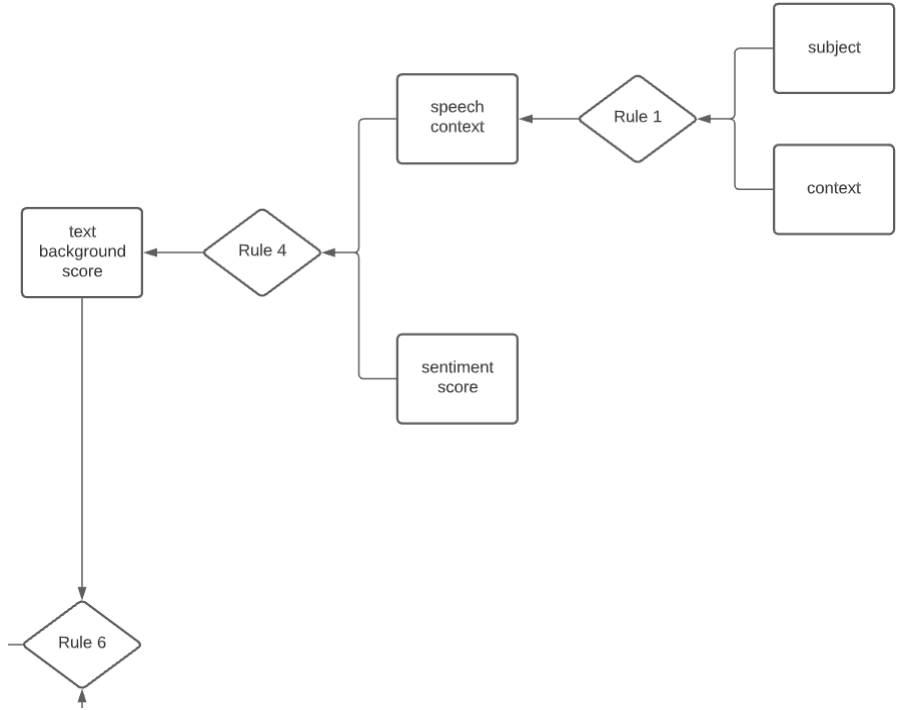

Figure 11. Text background related input facts network region.

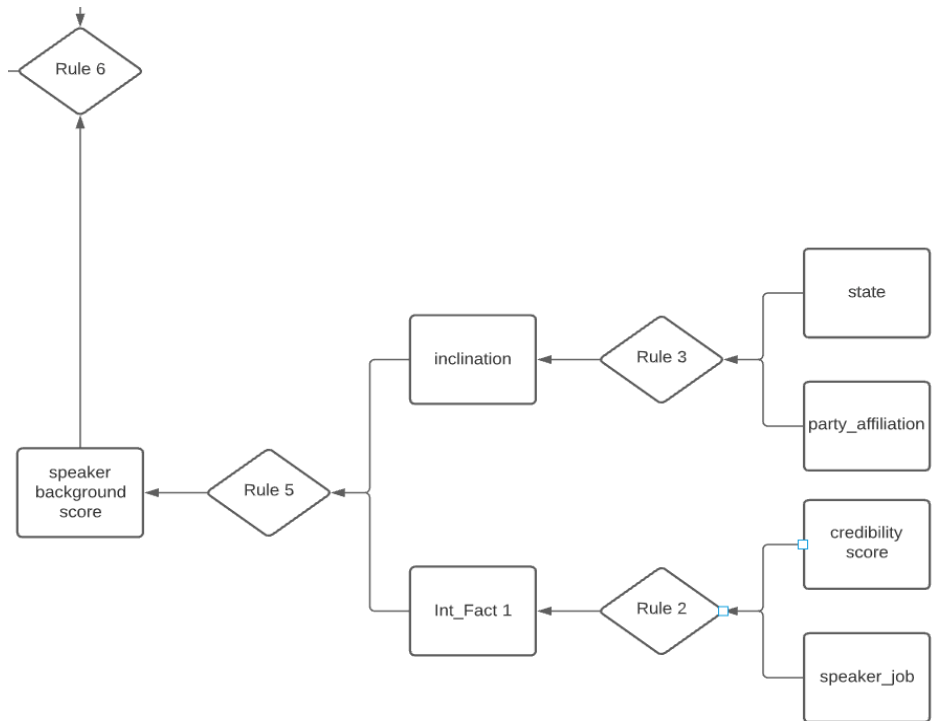

Figure 12. Speaker background related input facts network region.

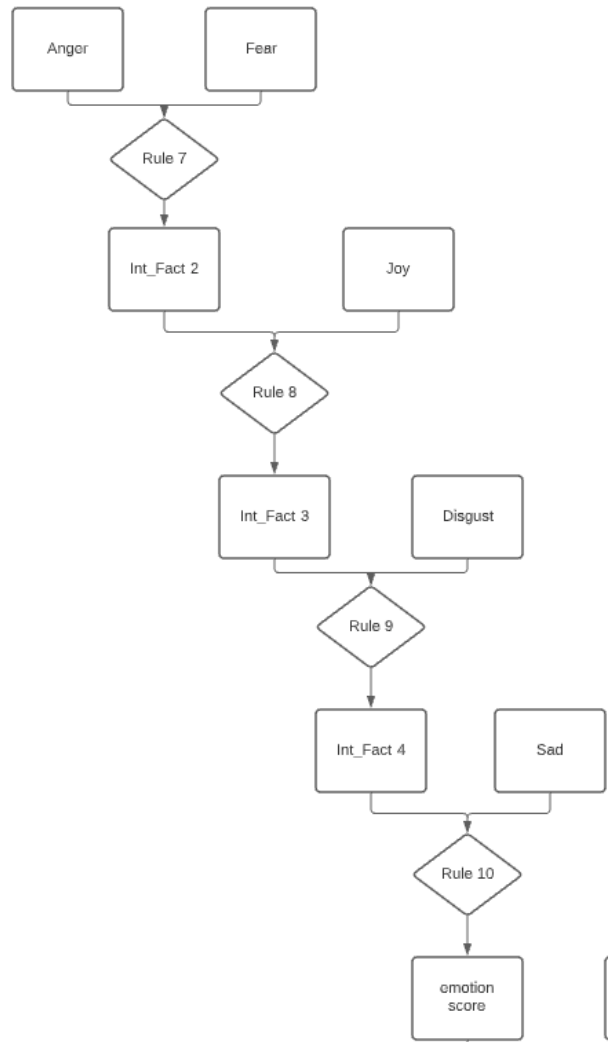

Figure 13. Emotion scores network region.

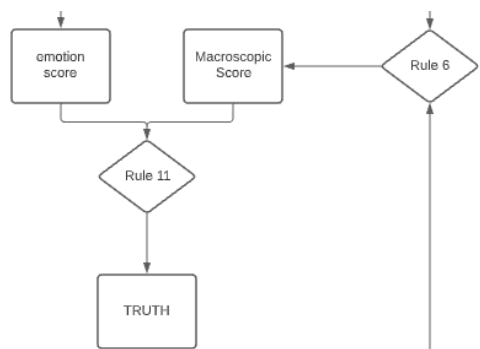

Figure 14. Facts that directly contribute to the output (TRUTH) value network region.

As the diagrams show, all the input facts are divided into three groups. Then, the text and speaker background related facts are combined into a macroscopic score, which is finally combined with the emotion score to produce the output value.

## 5. Study Comparison

The two studies that have been described in the previous sections have been conducted independently. However, there are some helpful observations that can be made by looking at the two of them together. This section compares and contrasts the studies in two key areas: their network designs and datasets.

*5.1. Networks designs*

While MLES is designed to be domain agnostic and is used across different domains, a key goal of using the system is to make the results of the machine learning process easily understandable and to prevent illogical and non-causal relationships from being learned by the system's training process. As the two studies discussed herein have shown, there is considerable flexibility in network design, while still achieving these goals.

A key benefit of MLES enjoyed by both studies is the ability to assess exactly which facts influence a given result. Thus, the fake news detection project – in addition to trying to develop a functional system – is also an attempt to understand which variables play a role in effectively detecting fake news. Similarly, the phishing project is – again, in addition to developing a functional system – an attempt to discover which variables are important in detecting phishing URLs. In both cases, the studies started from datasets that have been used for neural network-based classification. Thus, exact nature of the classification process and decision that was used (i.e., the internal workings of the neural network and the meaning of these workings) are unknown.

Both projects independently developed[3] their networks. In some ways, they produced similar results. The MLES design capabilities provided significant freedom in determining which associations between and encapsulations of facts to make, as well as flexibility in regards to the overall structure of the network.

The networks differ in their overall size and, of course, in their particular rule-fact composition. For the fake news identification study, the network has a smaller and more linear design. It is based on a dataset with a smaller number of fields. Some fields were pre-processed for importation into the MLES system.

The phishing study, on the other hand, has a larger network with a more interconnected structure. It started from a dataset with significantly more fields and somewhat reduced the number that are used through combining them prior to their importation into the MLES. Unlike the fake news detection study, the phishing study network includes certain facts as input facts to more than one rule.

This initial comparative study serves to document the design decisions made by these two projects as well as to demonstrate how MLES can be utilized. IT also serves to highlight the flexibility tha the MLES system provides with regards to supported network designs. However, in the longer term, the comparison of the results of these two studies will also inform future decision making in regards to whether the size and/or complexity of the MLES network plays a role (and, if so, to what extent it is impactful) in the effectiveness of system implementations. The analysis of this is a key area of planned future work.

*5.2. Datasets*

As discussed above, the two projects utilized datasets that were dissimilar interms of their composition and how they were utilized. While both were pre-processed, the phishing dataset had fields combnd to simplify the MLES network, prior to the data being imported into the system for training and testing.

---

[3] This is not meant to assert that there was no communications or collaboration between the projects, just that they didn't work together to build similar networks.

Despite these differences, and although the MLES does not have an explicit grouping feature, a similar grouping structure was utilized in both of the studies. The input facts in the phishing detection project are divided into two principal groups: URL properties and domain status. Similarly, the input facts in the fake news detection project are grouped into three groups based upon background information of the speech and speaker as well as the emotions revealed.

While the network structure has some similarities, the data used in each project differs in regards to how it is obtained. For the phishing project, the data is a direct property of the URL or website that can be reproducibly obtained. The fake news project, however, uses data that has significant subjectivity to it, in terms of truth/falsity status and other characteristics. The classification of emotions, for example, is highly subject (whether the decision is made by a person or a program such as IBM's Watson, which was used in making the Sentimental LIAR dataset [66]). These differences can potentially impact how effective classification can be and the effectiveness of various classification system techniques, which must deal with data noise, contradictions and errors which may prospectively result in degraded training and output data. As with the network structure, the analysis of the systems efficacy with regards to data characteristics will be a key area of future study.

## 6. Conclusion

This paper has discussed the use of the MLES system in two separate areas of study. It has demonstrated that the MLES system is able to support network designs for both areas, despite key differences in data and the network design approach taken by the researchers. It has, thus, demonstrated the effectiveness of the MLES system for the domains – future work will serve to characterize system effectiveness for the processing the particular datasets and evaluate the results of this processing.

Although phishing site identification and fake news detection are not new fields, minimal work in both areas has been done using explainable (much less defensible) machine learning systems. The phishing study explored the relationship between the various data factors, including symbol frequency, URL section lengths, and domain statistics such as registration duration and SSL support, which contribute to decision making. The fake news study, similarly, explored how various emotions, the context in which the statements are made, and the background of the speakers influence the likelihood of an article being intentionally deceptive 'fake news' content.

This paper has also discussed how the MLES can be used with real-world data and shown how it can be used with existing datasets that were developed for use with neural network techniques. Issues with traditional neural networks, which MLES is responsive to, have been discussed. Fundamentally, the goal of these two studies, as well as the MLES in general, is to make highly accurate decisions that system users and impacted individuals can understand and rely upon, due to the guarantees made by the MLES model of using a training-inviolate network.

In addition to seeking to answer their own research questions, these two studies also serve to demonstrate that the MLES can be applied to a wide range of areas of study and support varied different network designs. The fake news detection study utilizes a simple and linear network while the phishing websites detection study utilizes a more complex network. In the first study, facts are used once, while the second study utilizes input facts with multiple rules. The two studies have also demonstrated that the MLES can support multiple different types of data via simple pre-processing steps. The phishing study, for example, utilized data that included combinations of numbers, letters, and symbols. The fake news study, alternately, used data comprised of words and sentences.

Both studies have key areas of identified future work. In terms of network design, the fake news detection study plans to assess other datasets with additional (and different) variables. As part of this, different network designs may be evaluated, both in regards to new and existing data fields.

Both projects are undergoing the training and refinement using the MLES system to determine how to best support their application areas. Not only will this process produce functional systems for the two domain areas, it will also inform future development of the MLES system itself.

Both projects share a goal of helping to reduce the labor required for the detection of harmful conditions. They seek to prevent users from falling victim to their respective maladies. By doing this autonomously, they allow users to be protected across the internet – instead of just using (or only benefiting from the system when using) pre-screened websites and content. Of course, these benefits only accrue if users actually use the protection systems. Thus, the MLES system serves an important role of driving trust in warning (and potentially blocking) decision making by making it understandable and guaranteeing that only causal factors are considered in making restrictive decisions or recommendations.

**Acknowledgements**

This work has been partially supported by the U.S. National Science Foundation (NSF Award # 1757659). Facilities and other resources have been provided by the NDSU Department of Computer Science. Thanks is given to Logan Brown and Reid Pezewski for their feedback on network creation and a draft of this manuscript.

**References**

[1] "Neural networks for control," *Proc. 1999 Am. Control Conf. (Cat. No. 99CH36251)*, pp. 1642–1656 vol.3, 1999, doi: 10.1109/ACC.1999.786109.
[2] S. S. Keerthi, S. K. Shevade, C. Bhattacharyya, and K. R. K. Murthy, "Improvements to Platt's SMO algorithm for SVM classifier design," *Neural Comput.*, vol. 13, no. 3, pp. 637–649, Mar. 2001, doi: 10.1162/089976601300014493.
[3] G. H. John and P. Langley, "Estimating Continuous Distributions in Bayesian Classifiers," *undefined*, 1995.
[4] J. R. Quinlan, "C4.5: Programs for Machine Learning," *undefined*, 1992.
[5] D. W. Aha, D. Kibler, and M. K. Albert, "Instance-Based Learning Algorithms," *Mach. Learn.*, vol. 6, no. 1, pp. 37–66, 1991, doi: 10.1023/A:1022689900470.
[6] J. Han and M. Kamber, "Data Mining: Concepts and Techniques," *undefined*, 2000.
[7] K. Krishna and M. N. Murty, "Genetic K-means algorithm," *IEEE Trans. Syst. Man, Cybern. Part B Cybern.*, vol. 29, no. 3, pp. 433–439, 1999, doi: 10.1109/3477.764879.
[8] H. S. Park and C. H. Jun, "A simple and fast algorithm for K-medoids clustering," *Expert Syst. Appl.*, vol. 36, no. 2, pp. 3336–3341, Mar. 2009, doi: 10.1016/J.ESWA.2008.01.039.
[9] S. R. Eddy, "What is a hidden Markov model?," *Nat. Biotechnol. 2004 2210*, vol. 22, no. 10, pp. 1315–1316, Oct. 2004, doi: 10.1038/nbt1004-1315.
[10] S. U. Noble, "Algorithms of Oppression," *Algorithms of Oppression*, Mar. 2019, doi: 10.2307/J.CTT1PWT9W5.
[11] A. Rahman, "Algorithms of oppression: How search engines reinforce racism:," *https://doi.org/10.1177/1461444819876115*, vol. 22, no. 3, pp. 575–577, Sep. 2019, doi: 10.1177/1461444819876115.
[12] H. Lebovits, "Automating Inequality: How High-Tech Tools Profile, Police, and Punish the Poor," *Public Integr.*, vol. 21, no. 4, pp. 448–452, Jul. 2019, doi: 10.1080/10999922.2018.1511671.
[13] T. Doyle, "Weapons of Math Destruction: How Big Data Increases Inequality and Threatens Democracy," *Inf. Soc.*, vol. 33, no. 5, pp. 301–302, Oct. 2017, doi:

[14] "Trump Twitter: Republicans and Democrats split over freedom of speech | Donald Trump | The Guardian." .
[15] J. Straub, "Expert system gradient descent style training: Development of a defensible artificial intelligence technique," *Knowledge-Based Syst.*, p. 107275, Jul. 2021, doi: 10.1016/j.knosys.2021.107275.
[16] M. Nadimpalli, "Artificial Intelligence Risks and Benefits," *Int. J. Innov. Res. Sci. Eng. Technol.*, vol. 3297, 2007, Accessed: Jun. 25, 2020. [Online]. Available: https://www.researchgate.net/publication/319321806.
[17] S. He *et al.*, "Game player strategy pattern recognition and how UCT algorithms apply pre-knowledge of player's strategy to improve opponent AI," in *2008 International Conference on Computational Intelligence for Modelling Control and Automation, CIMCA 2008*, 2008, pp. 1177–1181, doi: 10.1109/CIMCA.2008.82.
[18] A. Tosun, A. Bener, and R. Kale, "AI-Based Software Defect Predictors: Applications and Benefits in a Case Study," Jul. 2010. Accessed: Jun. 25, 2020. [Online]. Available: https://www.aaai.org/ocs/index.php/IAAI/IAAI10/paper/view/1561.
[19] H. A. Yanco and J. Gips, "Preliminary investigation of a Semi-Autonomous Robotic Wheelchair Directed Through Electrodes," in *Proceedings of the Rehabilitation Engineering Society of North America Annual Conference*, Jun. 1997, pp. 414–416, Accessed: Jun. 25, 2020. [Online]. Available: www.cs.bc.edu/~gips/EagleEyes.
[20] Z. A. Baig, M. Baqer, and A. I. Khan, "A pattern recognition scheme for Distributed Denial of Service (DDoS) attacks in wireless sensor networks," in *Proceedings - International Conference on Pattern Recognition*, 2006, vol. 3, pp. 1050–1054, doi: 10.1109/ICPR.2006.147.
[21] S. C. Jacobsen *et al.*, "Research Robots for Applications in Artificial Intelligence, Teleoperation and Entertainment," *Int. J. Rob. Res.*, vol. 23, no. 4–5, pp. 319–330, Apr. 2004, doi: 10.1177/0278364904042198.
[22] R. Caruana and A. Niculescu-Mizil, "An empirical comparison of supervised learning algorithms," in *ACM International Conference Proceeding Series*, 2006, vol. 148, pp. 161–168, doi: 10.1145/1143844.1143865.
[23] Y. Duan, X. Chen, R. Houthooft, J. Schulman, and P. Abbeel, "Benchmarking Deep Reinforcement Learning for Continuous Control," 2016, Accessed: Jun. 25, 2020. [Online]. Available: https://github.com/.
[24] G. Paliouras, C. Papatheodorou, V. Karkaletsis, and C. . Spyropoulos, "Discovering user communities on the Internet using unsupervised machine learning techniques," *Interact. Comput.*, vol. 14, no. 6, pp. 761–791, Dec. 2002, doi: 10.1016/S0953-5438(02)00015-2.
[25] R. Rojas, "The Backpropagation Algorithm," in *Neural Networks*, Berlin: Springer Berlin Heidelberg, 1996, pp. 149–182.
[26] R. Battiti, "Accelerated Backpropagation Learning: Two Optimization Methods," *Complex Syst.*, vol. 3, no. 4, pp. 331–342, 1989, Accessed: Feb. 22, 2021. [Online]. Available: https://www.complex-systems.com/abstracts/v03_i04_a02/.
[27] C. Aicher, N. J. Foti, and E. B. Fox, "Adaptively Truncating Backpropagation Through Time to Control Gradient Bias," in *Proceedings of The 35th Uncertainty in Artificial Intelligence Conference*, Aug. 2020, pp. 799–808, Accessed: Feb. 22, 2021. [Online]. Available: http://proceedings.mlr.press/v115/aicher20a.html.
[28] L. Chizat and F. Bach, "Implicit Bias of Gradient Descent for Wide Two-layer Neural Networks Trained with the Logistic Loss," *Proc. Mach. Learn. Res.*, vol. 125, pp. 1–34, Jul. 2020, Accessed: Feb. 22, 2021. [Online]. Available: http://proceedings.mlr.press/v125/chizat20a.html.
[29] P. Zhao, P. Y. Chen, S. Wang, and X. Lin, "Towards query-efficient black-box adversary with zeroth-order natural gradient descent," *arXiv*, vol. 34, no. 04. arXiv, pp. 6909–6916, Feb. 18, 2020, doi: 10.1609/aaai.v34i04.6173.
[30] Z. Wu, Q. Ling, T. Chen, and G. B. Giannakis, "Federated Variance-Reduced Stochastic Gradient

Descent with Robustness to Byzantine Attacks," *IEEE Trans. Signal Process.*, vol. 68, pp. 4583–4596, 2020, doi: 10.1109/TSP.2020.3012952.

[31] J. N. D. Gupta and R. S. Sexton, "Comparing backpropagation with a genetic algorithm for neural network training," *Omega*, vol. 27, no. 6, pp. 679–684, Dec. 1999, doi: 10.1016/S0305-0483(99)00027-4.

[32] A. Saffaran, M. Azadi Moghaddam, and F. Kolahan, "Optimization of backpropagation neural network-based models in EDM process using particle swarm optimization and simulated annealing algorithms," *J. Brazilian Soc. Mech. Sci. Eng.*, vol. 42, no. 1, p. 73, Jan. 2020, doi: 10.1007/s40430-019-2149-1.

[33] D. Waterman, *A guide to expert systems*. Reading, MA: Addison-Wesley Pub. Co., 1986.

[34] V. Zwass, "Expert system," *Britannica*, Feb. 10, 2016. https://www.britannica.com/technology/expert-system (accessed Feb. 24, 2021).

[35] R. K. Lindsay, B. G. Buchanan, E. A. Feigenbaum, and J. Lederberg, "DENDRAL: A case study of the first expert system for scientific hypothesis formation," *Artif. Intell.*, vol. 61, no. 2, pp. 209–261, Jun. 1993, doi: 10.1016/0004-3702(93)90068-M.

[36] L. A. Zadeh, "Fuzzy sets," *Inf. Control*, vol. 8, no. 3, pp. 338–353, Jun. 1965, doi: 10.1016/S0019-9958(65)90241-X.

[37] J. M. Renders and J. M. Themlin, "Optimization of Fuzzy Expert Systems Using Genetic Algorithms and Neural Networks," *IEEE Trans. Fuzzy Syst.*, vol. 3, no. 3, pp. 300–312, 1995, doi: 10.1109/91.413235.

[38] S. Mitra and S. K. Pal, "Neuro-fuzzy expert systems: Relevance, features and methodologies," *IETE J. Res.*, vol. 42, no. 4–5, pp. 335–347, 1996, doi: 10.1080/03772063.1996.11415939.

[39] E. Styvaktakis, M. H. J. Bollen, and I. Y. H. Gu, "Expert system for classification and analysis of power system events," *IEEE Trans. Power Deliv.*, vol. 17, no. 2, pp. 423–428, 2002.

[40] J. M. McKinion and H. E. Lemmon, "Expert systems for agriculture," *Comput. Electron. Agric.*, vol. 1, no. 1, pp. 31–40, Oct. 1985, doi: 10.1016/0168-1699(85)90004-3.

[41] M. Kuehn, J. Estad, J. Straub, T. Stokke, and S. Kerlin, "An expert system for the prediction of student performance in an initial computer science course," 2017, doi: 10.1109/EIT.2017.8053321.

[42] O. Arsene, I. Dumitrache, and I. Mihu, "Expert system for medicine diagnosis using software agents," *Expert Syst. Appl.*, vol. 42, no. 4, pp. 1825–1834, 2015.

[43] B. Abu-Nasser, "Medical Expert Systems Survey," *Int. J. Eng. Inf. Syst.*, vol. 1, no. 7, pp. 218–224, Sep. 2017, Accessed: Jan. 17, 2021. [Online]. Available: https://papers.ssrn.com/sol3/papers.cfm?abstract_id=3082734.

[44] A. Chohra, A. Farah, and M. Belloucif, "Neuro-fuzzy expert system E_S_CO_V for the obstacle avoidance behavior of intelligent autonomous vehicles," *Adv. Robot.*, vol. 12, no. 6, pp. 629–649, Jan. 1997, doi: 10.1163/156855399X00045.

[45] W. A. Sandham, D. J. Hamilton, A. Japp, and K. Patterson, "Neural network and neuro-fuzzy systems for improving diabetes therapy," Nov. 2002, pp. 1438–1441, doi: 10.1109/iembs.1998.747154.

[46] E. P. Ephzibah and V. Sundarapandian, "A Neuro Fuzzy Expert System for Heart Disease Diagnosis," *Comput. Sci. Eng. An Int. J.*, vol. 2, no. 1, pp. 17–23, Feb. 2012, Accessed: Feb. 22, 2021. [Online]. Available: https://citeseerx.ist.psu.edu/viewdoc/download?doi=10.1.1.1052.1581&rep=rep1&type=pdf.

[47] S. Das, P. K. Ghosh, and S. Kar, "Hypertension diagnosis: A comparative study using fuzzy expert system and neuro fuzzy system," 2013, doi: 10.1109/FUZZ-IEEE.2013.6622434.

[48] B. A. Akinnuwesi, F. M. E. Uzoka, and A. O. Osamiluyi, "Neuro-Fuzzy Expert System for evaluating the performance of Distributed Software System Architecture," *Expert Syst. Appl.*, vol. 40, no. 9, pp. 3313–3327, Jul. 2013, doi: 10.1016/j.eswa.2012.12.039.

[49] J. Straub, "Machine learning performance validation and training using a 'perfect' expert system," *MethodsX*, vol. 8, p. 101477, Jan. 2021, doi: 10.1016/J.MEX.2021.101477.

[50] R. Dhamija, J. D. Tygar, and M. Hearst, "Why Phishing Works," *Proc. SIGCHI Conf. Hum. Factors Comput. Syst.*, 2006, doi: 10.1145/1124772.
[51] J. N., J. A., JakobssonMarkus, and MenczerFilippo, "Social phishing," *Commun. ACM*, vol. 50, no. 10, pp. 94–100, Oct. 2007, doi: 10.1145/1290958.1290968.
[52] "PhishTank | Join the fight against phishing." .
[53] C. Whittaker, B. Ryner, and M. Nazif, "Large-Scale Automatic Classification of Phishing Pages."
[54] G. Vrbančič, I. Fister, and V. Podgorelec, "Parameter Setting for Deep Neural Networks using Swarm Intelligence on Phishing Websites Classification," *Int. J. Arti-ficial Intell. Tools*, vol. 28, no. 6, pp. 1–28, 2019, doi: 10.1142/S021821301960008X.
[55] G. Vrbančič, I. Fister, and V. Podgorelec, "Datasets for phishing websites detection," *Data Br.*, vol. 33, p. 106438, Dec. 2020, doi: 10.1016/J.DIB.2020.106438.
[56] J. Gillin, "How Pizzagate went from fake news to a real problem for a D.C. business," *PolitiFact*, Dec. 05, 2016.
[57] H. Allcott and M. Gentzkow, "Social Media and Fake News in the 2016 Election," *J. Econ. Perspect.*, vol. 31, no. 2, 2017, doi: 10.1257/jep.31.2.211.
[58] M. Spradling, J. Straub, and J. Strong, "Protection from 'Fake News': The Need for Descriptive Factual Labeling for Online Content," *Futur. Internet*, vol. 13, no. 6, p. 142, May 2021, doi: 10.3390/fi13060142.
[59] "How Trump's 'fake news' rhetoric has gotten out of control - CNNPolitics." https://www.cnn.com/2019/06/11/politics/enemy-of-the-people-jim-acosta-donald-trump/index.html (accessed Feb. 01, 2020).
[60] "Tenn. lawmaker introduces resolution calling CNN, Washington Post 'fake news.'" https://www.wkyt.com/content/news/Tenn-lawmaker-introduces-resolution-calling-CNN-Washington-Post-fake-news-567446771.html (accessed Feb. 01, 2020).
[61] M. Duradoni, S. Collodi, S. C. Perfumi, and A. Guazzini, "Reviewing Stranger on the Internet: The Role of Identifiability through 'Reputation' in Online Decision Making," *Futur. Internet*, vol. 13, no. 5, p. 110, Apr. 2021, doi: 10.3390/fi13050110.
[62] "More Americans Are Getting Their News From Social Media." https://www.forbes.com/sites/petersuciu/2019/10/11/more-americans-are-getting-their-news-from-social-media/#589ec4d43e17 (accessed Feb. 01, 2020).
[63] C. Silverman and J. Singer-Vine, "Most Americans Who See Fake News Believe It, New Survey Says," *BuzzFeed News*, Dec. 06, 2016.
[64] J. T. Hancock, M. T. Woodworth, and S. Porter, "Hungry like the wolf: A word-pattern analysis of the language of psychopaths," *Leg. Criminol. Psychol.*, vol. 18, no. 1, pp. 102–114, Feb. 2013, doi: 10.1111/j.2044-8333.2011.02025.x.
[65] V. L. Rubin and T. Lukoianova, "Truth and deception at the rhetorical structure level," *J. Assoc. Inf. Sci. Technol.*, vol. 66, no. 5, pp. 905–917, May 2015, doi: 10.1002/asi.23216.
[66] B. Upadhayay and V. Behzadan, "Sentimental LIAR: Extended Corpus and Deep Learning Models for Fake Claim Classification," Nov. 2020, doi: 10.1109/ISI49825.2020.9280528.
[67] "Watson Natural Language Understanding | IBM." .
[68] W. Y. Wang, "'Liar, Liar Pants on Fire': A New Benchmark Dataset for Fake News Detection," pp. 422–426, 2017.